\documentclass[showpacs,preprintnumbers,prd,amsmath,amssymb]{revtex4}
\usepackage{amsmath}
\usepackage{graphicx}
\usepackage{subfig}
\usepackage{hyperref}
\usepackage{amssymb}
\usepackage[english]{babel}
\usepackage{epsfig}
\usepackage{wasysym}

\begin{document}

\title{Comment on ``Generic rotating regular black holes in general relativity coupled to
non-linear electrodynamics''\\}

\author{ Manuel E. Rodrigues$^{(a,b)}$\footnote{E-mail address: 
esialg@gmail.com} and  Ednaldo L. B. Junior$^{(b,c)}$\footnote{E-mail 
address: ednaldobarrosjr@gmail.com}}
\affiliation{$^{(a)}$Faculdade de Ci\^{e}ncias Exatas e Tecnologia, 
Universidade Federal do Par\'{a}\\
Campus Universit\'{a}rio de Abaetetuba, 68440-000, Abaetetuba, Par\'{a}, 
Brazil\\
$^{(b)}$Faculdade de F\'{\i}sica, Programa de P\'os-Gradua\c{c}\~ao em 
F\'isica, Universidade Federal do 
 Par\'{a}, 66075-110, Bel\'{e}m, Par\'{a}, Brazil\\
$^{(c)}$Faculdade de Engenharia da Computa\c{c}\~{a}o, 
Universidade Federal do Par\'{a}, Campus Universit\'{a}rio de Tucuru\'{\i}, 
68464-000, Tucuru\'{\i}, Par\'{a}, Brazil}


\begin{abstract}
We show that there is an inconsistency in the class of solutions obtained in Phys.
Rev. D {\bf 95}, 084037 (2017). This inconsistency is due to the approximate relation between
lagrangian density and its derivative for Non-Linear Electrodynamics. We present an algorithm to obtain new classes of solutions.
\end{abstract}

\date{\today}

\maketitle



\section{Demonstration}
\label{sec1}

The action of General Relativity coupled to Non-linear Electrodynamics (NED) is given by
\begin{eqnarray}
S=\int dx^4\sqrt{-g}\left[R+8\pi G\mathcal{L}_{NED}(F)\right]\,,\label{action}
\end{eqnarray}
where $R$ is the scalar curvature and $F=F_{\mu\nu}F^{\mu\nu}$, with $F_{\mu\nu}=\partial_{\mu}A_{\mu}-\partial_{\nu}A_{\mu}$ and $A_{\mu}$ the four-potential. The functional variation of the action\eqref{action} in relation to metric and the four-potential $A_{\mu}$ results in
\begin{eqnarray}
&&R_{\mu\nu}-\frac{1}{2}g_{\mu\nu}R=8\pi GT_{\mu\nu},\label{eq}\\
&&\nabla_{\mu}\left[F^{\mu\nu}\mathcal{L}_{F}\right]=0,\label{eqm2}
\end{eqnarray} 
where
\begin{eqnarray}
&&T_{\mu\nu}=g_{\mu\nu}\mathcal{L}_{NED}-\mathcal{L}_FF_{\mu}^{\;\;\alpha}F_{\nu\alpha},\\
&&\mathcal{L}_F=\frac{\partial \mathcal{L}_{NED}}{\partial F}.\label{eq1}
\end{eqnarray}
Constructing a regular metric with axial symmetry through NEWMAN-JANIS algorithm we have \cite{toshmatov}
\begin{eqnarray}
ds^2=\left(1-\frac{2\rho r}{\Sigma}\right)dt^2-\frac{\Sigma}{\Delta}dr^2+2a\sin^2\theta\frac{2\rho r}{\Sigma}dtd\phi -\Sigma d\theta-\sin^2\theta\frac{(r^2+a^2)^2-a^2\Delta\sin^2\theta}{\Sigma}d\phi^2\label{ele}.
\end{eqnarray}
with $\Sigma=r^2+a^2\cos^2\theta,\Delta=r^2-2r\rho+a^2$ and $\rho(r)$ the mass function, where $\lim_{r\rightarrow +\infty}\rho(r)=M_{ADM}$ with $M_{ADM}$ the ADM mass.
\par 
For a space-time with axial symmetry, the Maxwell tensor obeys the following equation $\mathbb{L}_{K^{\alpha}}F^{\mu\nu}\equiv 0$, where $\mathbb{L}$ represents the Lie's derivative and $K^{\alpha}$ the Killing vectors of axial symmetry. Through this symmetry we have the following non-zero components for the Maxwell tensor \cite {dymnikova1} $F_{01}(t,r,\theta),F_{02}(t,r,\theta),F_{13}(t,r,\theta)$ and $F_{23}(t,r,\theta)$. As the solution must be stationary and coming from a NEWMAN-JANIS algorithm, the following constraint can be made for the four-potential components \cite{erbin}
\begin{eqnarray}
A_{\mu}=\frac{Q a\cos\theta}{\Sigma}\delta^{0}_{\mu}-\frac{Q(r^2+a^2)\cos\theta}{\Sigma}\delta^{3}_{\mu}.\label{A}
\end{eqnarray} 
Now, according to \cite{toshmatov}, calculating the Maxwell tensor, the scalar $F$ is given by
\begin{eqnarray}
F=\frac{Q^2}{4\Sigma^4}[a^4(3-\cos 4\theta)+4(6a^2r^2+2r^4+a^2(a^2-6r^2)\cos 2\theta)]\label{F}.
\end{eqnarray}
Agin  according to \cite{toshmatov}, using \eqref{ele} and \eqref{A} in \eqref{eq}, solving the equations in terms to $\mathcal{L}_{NED}$ and $\mathcal{L}_F$ we have
\begin{eqnarray}
&&\mathcal{L}_{NED}=\frac{r^2}{2\Sigma^4}\Big[\left(15a^4-8a^2r^2+8r^4+4a^2(5a^2-2r^2)\cos 2\theta +5a^4\cos 4\theta\right)\rho^{\prime}+16a^2r\cos^2\theta\Sigma\rho^{\prime\prime}\Big]\label{L}\\
&&\mathcal{L}_F=\frac{2(r^2-a^2\cos^2\theta)\rho^{\prime}-r\Sigma\rho^{\prime\prime}}{2Q^2}.\label{LF}
\end{eqnarray}

The central question here is that solving like equations of motion \eqref{eq}, the Lagrangian density and its derivative will not necessarily be related as in \eqref{eq1}. This should be imposed so that the solution is consistent with NED. Let's now check this property. We can rewrite \eqref{eq1} as
\begin{eqnarray}
\mathcal{L}_F-\left[\frac{\partial \mathcal{L}_{NED}}{\partial r}\left(\frac{\partial F}{\partial r}\right)^{-1}+\frac{\partial \mathcal{L}_{NED}}{\partial \theta}\left(\frac{\partial F}{\partial \theta}\right)^{-1}\right]=0\label{eqcons}.
\end{eqnarray}
Taking \eqref{F}, \eqref{L} and \eqref{LF}, the consistency equation \eqref{eqcons}  becomes
\begin{eqnarray}
&&\mathcal{L}_F-\left[\frac{\partial \mathcal{L}_{NED}}{\partial r}\left(\frac{\partial F}{\partial r}\right)^{-1}+\frac{\partial \mathcal{L}_{NED}}{\partial \theta}\left(\frac{\partial F}{\partial \theta}\right)^{-1}\right]=\frac{1}{4Q^2}\Big\{2\left[2(r^2-a^2\cos^2\theta)\rho^{\prime}-r(r^2+a^2\cos^2\theta)\rho^{\prime\prime}\right]+\nonumber\\
&&+\frac{16r^4\left[(6a^2-4r^2+6a^2\cos 2\theta)\rho^{\prime}+r(a^2+2r^2+a^2\cos 2\theta)\rho^{\prime\prime}\right]}{a^4+16a^2r^2+16r^4+2a^2(a^2-2r^2)\cos 2\theta}+\nonumber\\
&&-\Big[-2(a^2-2r^2+a^2\cos 2\theta)^2(5a^2+2r^2+5a^2\cos 2\theta)\rho^{\prime}-\frac{1}{2}r(a^2+2r^2+a^2\cos 2\theta)((33a^4+8r^4+44a^4\cos 2\theta +\nonumber\\
&&+11a^4\cos 4\theta)\rho^{\prime\prime}+8a^2r\cos^2\theta(a^2+2r^2+a^2\cos 2\theta)\rho^{\prime\prime\prime})\Big][a^2(4r^2+2(a^2-4r^2)\cos 2\theta +a^2\cos 4\theta)]^{-1}\Big\}.\label{eqcons1}
\end{eqnarray} 
Now using the model of mass function (30) on \cite{toshmatov}
\begin{eqnarray}
\rho(r)=M+\frac{\alpha^{-1}q^3r^{\mu}}{(r^{\nu}+q^{\nu})^{\mu/\nu}}\label{mass},
\end{eqnarray}
we can explicitly verify that the consistency equation \eqref{eqcons1} is not satisfied as in \eqref{eqcons}. Therefore any solution coming from  NEWMAN-JANIS algorithm, whose Lagrangean density and its derivative are \eqref{L} and \eqref{LF}, with mass function give by \eqref{mass} is inconsistent. Some spherically symmetrical models of regular black holes have consistently used this relationship  \cite{manuel} and \cite{fan}.
\par 
Despite, the discussed inconsistency in Eq. \eqref{eqcons}, one could see that the derived solution
in Ref. \cite{toshmatov} is one of the most possible solutions, since it gives the correct results in
non rotating limit as in Ref. \cite{fan}. Furthermore, the numerical calculations show that the
difference between two parts of Eq. \eqref{eqcons} is comparably small. The solution obtained in
Ref. \cite{toshmatov} can be considered as an approximate solution with the comparatively small incorrectness in the behaviour of the electromagnetic field. Moreover, there is well-accepted fact that similar notion can be related to many other published solutions in Non-Linear Electrodynamics.
\par
We conclude this section by emphasizing that the class of rotating regular black holes solutions  obtained in \cite{toshmatov} does not fulfil the whole set of field equations in \eqref{eq} and \eqref{eqm2}, which will be better explained in the next section.


\section{Possibility to new solutions}
\label{sec2}

The NEWMAN-JANIS algorithm can present some pathology \cite{hansen} in some cases, but its generalization does not appear to present these pathologies \cite{mustapha}. The authors to \cite{toshmatov} has used the generalization of this algorithm. The algorithm itself seems to show no inconsistency, but the way in which it was used by the authors of \cite{toshmatov}, it was not correct. Let's comment this better now.
\par 
We will start this section by explaining why the authors of \cite{toshmatov} did not obtain an exact solution of the theory coming from the action \ref{action} with axial symmetry. First, they did not begin by integrating the modified Maxwell equation to NED, thus suggesting a solution to the potential in equation (25) de \cite{toshmatov}. Second, they attempted to solve Einstein's equations $G_{\mu\nu}=T_{\mu\nu}$, which are five equations with only three arbitrary functions $\{\rho(r),\mathcal{L},\mathcal{L}_F\}$. This eventually left some of the equations of motion (Einstein and Maxwell modified) not satisfied. 
\par 
Now let's point out a way to obtain exact solutions for this system. First we must integrate the modified Maxwell equations
\begin{eqnarray}
\nabla_{\mu}\left[\mathcal{L}_F F^{\mu\nu}\right]=0,\nabla_{\mu}\; ^{*}F^{\mu\nu}=0,\nabla_{[\mu} F_{\alpha\beta]}=0,^{*}F^{\mu\nu}=\frac{1}{2}\eta^{\mu\nu\alpha\beta}F_{\alpha\beta},\eta^{0123}=\frac{-1}{\sqrt{-g}}.
\end{eqnarray}
The non-zero Maxwell tensor components are $\{F_{10},F_{20},F_{13},F_{23}\}$, but two are dependent on the others \cite{D}, remaining only $\{F_{10},F_{20}\}$. These Maxwell tensor components may perhaps be integrated to obtain a solution that depends on the Lagrangian and its derivative $\{\mathcal{L},\mathcal{L}_F\}$. We could then insert them into the Einstein equations projected with the tetrads $G^{(a)}_{\;\;(b)}=T^{(a)}_{\;\;(b)}$, where we have only two equations for the case $T^{(2)}_{\;\;(2)}=T^{(3)}_{\;\;(3)}$. We can now solve the two equations to obtain $\{\mathcal{L},\mathcal{L}_F\}$, remember that there is a constraint equation $\mathcal{L}_F=\partial \mathcal{L}/\partial F$, which if soluble, then we have a new class of exact solutions.
\par 
The algorithm specified above to obtain new classes of solutions is not simple and it may be that the equation $\mathcal{L}_F=\partial \mathcal{L}/\partial F$ is highly non-linear in the mass function and does not present solution by known methods. So this should be a future work separate from this comment.


\end{document}